\documentclass[12pt]{iopart}
\usepackage{iopams}
\usepackage{graphicx}
\expandafter\let\csname equation*\endcsname\relax
\expandafter\let\csname endequation*\endcsname\relax 
\usepackage{amsmath,amssymb}
\usepackage[normalem]{ulem}
\usepackage{braket}
\usepackage{physics}
\usepackage{bbold}

\newcommand{\beq}{\begin{equation}}
\newcommand{\eeq}{\end{equation}}
\newcommand{\beqar}{\begin{eqnarray}}
\newcommand{\eeqar}{\end{eqnarray}}

{\large{}}
\global\long\def\ave#1{\langle#1\rangle}%
{\large\par}

{\large{}}
{\large\par}

{\large{}}
{\large\par}

{\large{}}
{\large\par}

{\large{}}
{\large\par}

{\large{}}
\global\long\def\ha{\hat{a}}%
{\large\par}

{\large{}}
{\large\par}

{\large{}}
\global\long\def\hH{\hat{H}}%
{\large\par}

{\large{}}
\global\long\def\hH{\hat{H}}%
{\large\par}

{\large{}}
\global\long\def\had{\hat{a}^{\dagger}}%
{\large\par}

\usepackage{tikz}
\usetikzlibrary{shadows}
\usetikzlibrary{arrows.meta,calc,decorations.markings,math,positioning,arrows,automata}
\usetikzlibrary{shapes.geometric}
\usepgflibrary{arrows.meta}
\tikzset{>=latex}
\tikzset{->-/.style={decoration={
  markings,
  mark=at position .5 with {\arrow{>}}},postaction={decorate}}}
  
\tikzset{-<-/.style={decoration={
  markings,
  mark=at position .5 with {\arrow{<}}},
  postaction={decorate}}}
  
 \tikzset{arrow data/.style 2 args={%
     decoration={%
        markings,
        mark=at position #1 with \arrow{#2}},
        postaction=decorate}
     }%

\begin{document}
\title{Single-atom heat engine as a sensitive thermal probe}
\author{Amikam Levy$^{1,2,\dagger}$, Moritz G{\"o}b$^{3,\dagger}$, Bo Deng$^{3}$, Kilian Singer$^{3}$, E. Torrontegui$^{4}$, Daqing Wang$^{3}$}

\address{$^{1}$ Department of Chemistry, University of California Berkeley, Berkeley, California 94720, United States}
\address{$^{2}$ The Sackler Center for Computational Molecular Science, Tel Aviv University, Tel Aviv 69978, Israel}
\address{$^{3}$ Experimentalphysik I, Universit\"at Kassel, Heinrich-Plett-Str. 40, 34132 Kassel, Germany}
\address{$^{4}$ Instituto de F\'{\i}sica Fundamental IFF-CSIC, Calle Serrano 113b, 28006 Madrid, Spain}
\address{$^{\dagger}$ these authors contributed equally}

\ead{amikamlevy@gmail.com, eriktorrontegui@gmail.com, ks@uni-kassel.de, daqing.wang@uni-kassel.de}

\begin{abstract}
We propose employing a quantum heat engine as a sensitive probe for thermal baths.
In particular, we study a single-atom Otto engine operating in an open thermodynamic cycle. Owing to its cyclic nature, the engine is capable of translating small temperature differences between two baths into a macroscopic oscillation in a flywheel. We present analytical and numerical modeling of the quantum dynamics of the engine and estimate it to be capable of detecting temperature differences as small as 2\,$\mu$K. This sensitivity can be further improved by utilizing quantum resources such as squeezing of the ion motion. The proposed scheme does not require quantum state initialization and is able to detect small temperature differences even at high base temperatures.
\end{abstract}

\section{Introduction}
Preparation and manipulation of quantum systems require the ability to detect physical properties of their environment with high precision. Many of the environment variables cannot be measured directly, as they are not proper quantum observables and must instead be inferred from indirect measurements. The temperature of a system is such a property; it can only be recovered from measurements of related observables. In recent years, the field of quantum thermometry~\cite{mehboudi2019thermometry,de2018quantum,mann2014quantum} has advanced tremendously due to its broad implications for quantum technologies~\cite{yue2012nanoscale,kucsko2013nanometre,neumann2013high,correa2015individual,raitz2015experimental}. Using a small quantum system as a probe to measure the temperature of a bath has the advantage of minimally perturbing the state of the bath. Furthermore, it was suggested that quantum phenomena can be employed to enhance the precision of thermometry~\cite{toyli2013fluorescence,jevtic2015single,mancino2017quantum}.

The emerging field of thermodynamics in the quantum regime~\cite{levy2014rev,kosloff13,janet2016quantum,alicki2018introduction} sheds new light on energy exchange processes between small quantum systems and their environment. Manipulating the interactions between the quantum system and the environment in a structured manner reveals simple laws that relate properties of the environment to measurable observables of the quantum system. For example, the power output of a quantum heat engine operating between two baths strongly depends on the temperatures of the baths.
While theoretical and experimental studies of quantum thermal devices have focused mainly on the efficiency, power output, cooling rates, etc. of the devices~\cite{levy2014rev,levy212,ghosh2018thermodynamic,correa2014optimal,allahverdyan2013carnot,gelbwaser2015thermodynamics} and their relation to quantum effects~\cite{levy2018book,uzdin15,klatzow2019experimental}, we wish to employ these results for parameter estimations of the baths. A similar idea was recently proposed in Ref.~\cite{hofer2017quantum} based on a quantum refrigerator model.  

\begin{figure}[t]
\centering
\includegraphics[width=14.5cm]{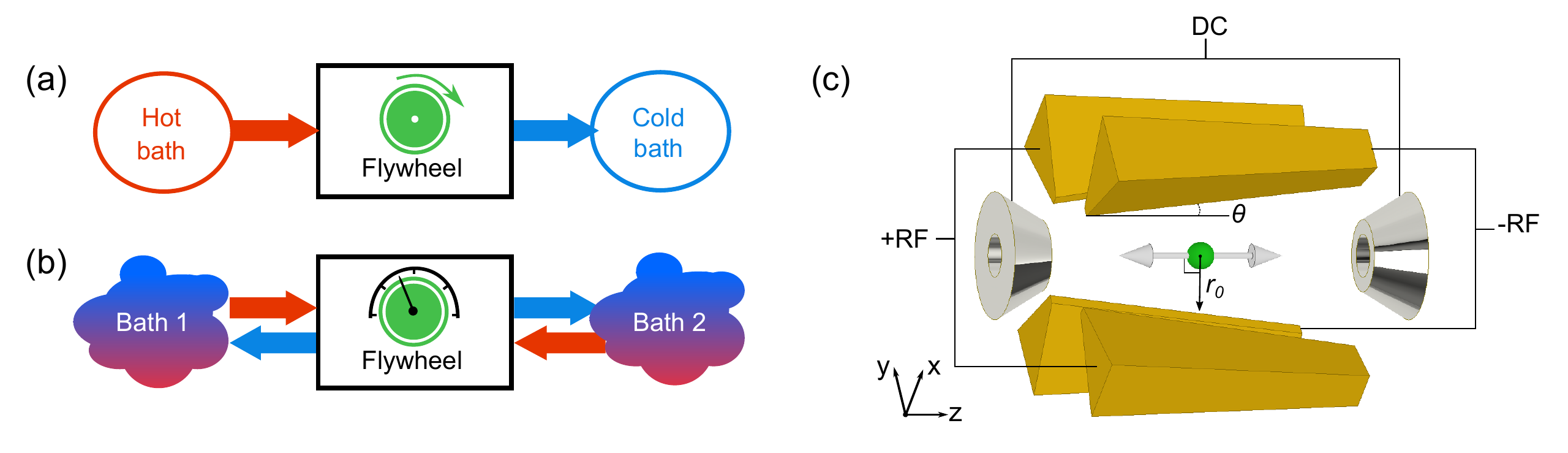}
\caption{(a) A heat engine takes in heat from a hot thermal bath, converts part of the thermal energy into mechanical work and releases the rest to a cold bath. The work can be stored in a flywheel. (b) A calibrated heat engine can be employed to measure the temperature difference between two baths by monitoring the energy in the flywheel. (c) A modified linear Paul trap with four tapered blade electrodes facilitating the operation of a single-ion heat engine.}
\label{Fig:scheme}
\end{figure}

The central idea of this work is to employ a heat engine operating in open thermodynamic cycles to evaluate the temperature difference between two thermal baths. As shown in Fig.\,\ref{Fig:scheme}a, a common heat engine takes in thermal energy from a hot bath, converts part of the energy into work and releases the rest to a cold bath. A flywheel stores the work output in its oscillatory motion and can potentially drive an external load~\cite{levy16,von2019spin}. In cases where the coupling between the bath and the flywheel can be quantitatively modelled, one can evaluate the properties of the baths by performing measurements on the flywheel (see Fig.\,\ref{Fig:scheme}b). We consider, in particular, the experimental setup of the single-atom heat engine~\cite{rossnagel2016} but with a non-damped flywheel. In this case, the energy in the flywheel grows quadratically with the number of engine cycles and avoids the possible exponential growth of the fluctuations~\cite{levy16}.

Operating the engine in the quantum regime permits the utilization of quantum resources such as squeezing of the working medium. Theoretical and experimental studies of quantum engines operating between non-thermal baths predict that the efficiency and power output may exceed the bounds set by thermal engines~\cite{rossnagel2014,niedenzu2018quantum,klaers2017squeezed}. While these results might not be entirely surprising, as an additional source of energy can be exploited, we show that squeezing the working medium after the thermalizing strokes can significantly amplify the energy stored in the flywheel. This enhances the sensitivity of the thermal probe, enabling the detection of very small temperature differences which otherwise could not be resolved in experiments.

The manuscript is structured as follows. In Sect.\,\ref{sec:AnalyticModel}, we present the single-ion heat engine model and analyze its dynamics. In Sect.\,\ref{sec:sensor}, we introduce the measurement protocol and estimate the sensitivity based on experimental parameters. Finally, in Sect.\,\ref{sec:discussions} we discuss the possible extensions and limitations of this scheme.

\section{Dynamics of the single-ion heat engine}
\label{sec:AnalyticModel}
We start by introducing the experimental setup and a theoretical model which describes the dynamics of the working medium and the flywheel of the engine. 

\subsection{Equations of motion}
\label{eqs_motion}
As depicted in Fig.\,\ref{Fig:scheme}c, an atomic ion is trapped in a Paul trap consisting of four blade-shaped electrodes and two endcap-electrodes\,\cite{rossnagel2016}. In contrast to a conventional Paul trap, the blade electrodes are tilted with respect to the axial symmetry axis by an angle~$\theta$. The four blade-electrodes are driven by radio-frequency (RF) signals and the two endcap-electrodes are biased with positive DC voltages. The ponderomotive potential formed by this trap can be approximated as harmonic in the radial ($x$, $y$) and axial ($z$) directions with a variation of the radial trapping frequencies in the axial direction:
\begin{equation}
\label{freq}
\omega_{x,\,y}(z)=\frac{\omega_{x0,\,y0}}{(1+\tan{\theta}\cdot z/r_0)^2}\,,
\end{equation}
with $r_0$ the radial distance of the ion to the blade electrodes and $\omega_{x0,\,y0}$ the radial trapping frequencies at $z=0$. In an experiment, a small anisotropy of the radial potential is normally present, which lifts the degeneracy of $\omega_{x0,\,y0}$. The confinement in the axial direction is much weaker compared to that of the radial directions, signified by a smaller axial trapping frequency $\omega_z\approx\omega_{x0,\,y0}/10$.

The Hamiltonian describing the motional state of an ion in this potential can be written as
\begin{equation}
    \hat H=\frac{\hat p ^2_x}{2m}+\frac{1}{2}m\omega_x^2\hat x^2+\frac{\hat p ^2_y}{2m}+\frac{1}{2}m\omega_y^2\hat y^2+\frac{\hat p ^2_z}{2m}+\frac{1}{2}m\omega_z^2\hat z^2,
\end{equation}
with $m$ the mass of the ion, $\hat x$, $\hat y$, $\hat z$ and $\hat p_x$, $\hat p_y$, $\hat p_z$ the position and momentum operators in the corresponding directions. Viewed as a heat engine, the radial states represent the working medium which thermalizes with external hot and cold baths periodically. The engine drives and amplifies a coherent oscillation of the ion in the axial direction. The axial oscillator thus serves as a flywheel that stores the energy output. In open-cycle operations, the flywheel is not actively damped. As we will show in later sections, the energy stored in the flywheel grows quadratically with the number of engine cycles. In our analysis, we will treat the large-amplitude coherent oscillation in the axial direction as a classical oscillator, while keeping a quantum mechanical description for the radial states. The Hamiltonian of the system can then be rewritten as
\begin{eqnarray}
\label{Eq:H2}\nonumber
    \hat H &=\hbar\omega_{x0}(\hat{a}_x^{\dagger}\hat{a}_x+\frac{1}{2})+\hbar\omega_{y0}(\hat{a}_y^{\dagger}\hat{a}_y+\frac{1}{2})\\
    &+\frac{1}{2}m(\omega^2_{x0}\hat x^2+\omega^2_{y0}\hat y^2)\left(\frac{1}{\left(1+\gamma z\right)^4}-1\right)+\frac{p ^2_z}{2m}+\frac{1}{2}m\omega_z^2 z^2\,,
\end{eqnarray}
where $\hat{a}_x^{\dagger}$ ($\hat{a}_y^{\dagger}$) and $\hat{a}_x$ $(\hat{a}_y)$ represent the bosonic creation and annihilation operators of the phonons in $x$ ($y$) direction, respectively. The first two terms in Eq.\,(\ref{Eq:H2}) describe the energy stored in the working medium. The third term represents the coupling of the working medium to the flywheel, where we define $\gamma=\tan{\theta}/r_0$. The last two terms denote the kinetic and potential energy in the flywheel, where $p_z$ and $z$ are the classical momentum and position of the ion in the axial direction.

As indicated by Eq.\,(\ref{Eq:H2}), the two radial directions contribute to the Hamiltonian independently. For the sake of brevity, we consider only the $x$ direction in our further analysis, while assuming that the oscillator in the $y$ direction is maintained at a low and constant temperature. We denote the creation and annihilation operators in the $x$ direction as $\hat{a}^{\dagger}$ and $\hat{a}$.
The equations of motion for the quantum harmonic oscillator in the $x$ direction read
\begin{align}
\label{eq_radial}
\frac{d}{dt}\hat{X}(t) & =2\left(\hbar\omega_{x0}+2g(t)\right)\hat{Y}(t)\, \\
\frac{d}{dt}\hat{Y}(t) & =-2\left(\hbar\omega_{x0}+2g(t)\right)\hat{X}(t)-8g(t)\hat{N}(t)-4g(t)\,\nonumber \\
\frac{d}{dt}\hat{N}(t) & =-2g(t)\hat{Y}(t), \nonumber
\end{align}
where we define
\begin{align}
\label{eq_motion}
\hat{X}(t) & =  \left(\hat a^{\dag 2}(t)+\hat a^{2}(t)\right) \\
\hat{Y}(t) & =  i\left(\hat a^{\dag 2}(t)-\hat a^{2}(t)\right)\nonumber \\
\hat{N}(t) & =  \had(t)\ha(t)\nonumber \\
g(t) & =  \frac{\hbar\omega_{x0}}{4}\left(\frac{1}{[1+\gamma z(t)]^{4}}-1\right).\ \nonumber
\end{align}
The classical equations of motion for the flywheel can be written as
\begin{eqnarray}\nonumber
\dot{z}(t)&=&p_z(t)/m\\
\dot{p}_{z}(t)&=&\mathcal{F}(t)\,,
\label{Eq:eom1}
\end{eqnarray}
with $\mathcal{F}$ the force acting on the axial oscillator. In the mean-field approximation, $\mathcal{F}$ can be expressed as
\begin{eqnarray}\label{force}
    \mathcal{F}(t)&=&-\frac{\partial{\langle \hat H\rangle}}{\partial z}=-m\omega_z^2 z(t)+F(t)\,,
\end{eqnarray}
which is the sum of the restoring force of the harmonic potential and the force $F$ resulting from the radial-axial coupling
\begin{eqnarray}\label{radialforce}
    F(t)&=&\frac{\gamma\hbar\omega_{x0}{R}(t)}{\left(1+\gamma z(t)\right)^5}-\frac{\hbar\omega_{x0}}{4}\frac{\partial{R}}{\partial z}\left(\frac{1}{\left(1+\gamma z(t)\right)^4}-1\right).
\end{eqnarray}
Here, we define $R(t)=\ave{\left(\had(t)+\ha(t)\right)^{2}}$. We note that, one could alternatively assume the same temperature of the two radial modes, which would result in doubling of the force $F$. In typical experimental scenarios\,\cite{rossnagel2016}, the small axial displacement $z(t)\ll r_0 $ holds, leading to $\gamma z(t) \ll 1$. This allows the approximation
\begin{eqnarray}\label{force_approx}
    \mathcal{F}(t)&= &-m\omega_z^2 z(t)+ \gamma\hbar\omega_{x0}R(t),
    \label{Eq:axialforce}
\end{eqnarray}
with
\begin{align}
R(t) & = \ave{\hat{X}(t)+2\hat{N(t)}+1}\label{eq:second_moment}\\
& = 1+2N_{0}
 +X_{0}\cos\left(2\omega_{x0} t\right)
+\ Y_{0}\sin\left(2\omega_{x0} t\right), \nonumber
\end{align}
and $X_0$, $Y_0$, $N_0$ the expectation values of $\hat{X}(t)$, $\hat{Y}(t)$ and $\hat{N}(t)$ at $t=0$.

\subsection{Thermodynamic cycle}
The model can be cast to describe the dynamics of a four-stroke Otto engine \cite{Abah2012, Rezek2006, Torrontegui2013_b, Kosloff2017}. 
One engine cycle, as shown in {Figs.\,\ref{Fig:enginecycle}a,\,b, can be described as follows:
\textit{Hot isochore}~[A]: the radial trapping frequency is kept constant at $\omega_h$ and the working medium is in contact with a hot bath at temperature $T_h$. After a time of $\tau_h$, the working medium thermalizes at this temperature. The time scale of the interaction is much shorter than the half period of the axial oscillation $\tau_z=\pi/\omega_z$. During this stroke, the axial displacement of the ion and the change of the radial trapping frequency are negligible. \textit{Isentropic expansion}~[B]: the ion is isolated from the baths and evolves in the trapping potential. After a time of $\tau_z$, the radial trapping frequency changes from $\omega_h$ to $\omega_c$ due to the displacement in the axial direction. This trapping frequency change may be either adiabatic or
engineered using optimal control techniques \cite{Torrontegui2018} that also allow for the minimization of different sources of error in the variation of $\omega_x$ \cite{Levy2017, Levy2018}. 
The dynamics describing this process is unitary and the entropy of the working medium remains constant. 
\textit{Cold isochore}~[C]: the radial trapping frequency remains constant at $\omega_c$. 
The ion is coupled to a cold bath and its radial state thermalizes to the temperature $T_c$ ($T_c<T_h$) after time $\tau_c$ ($\tau_c\approx\tau_h\ll\tau_z$). \textit{Isentropic compression}~[D]: in the last stroke, the ion is again isolated from the baths and evolves isentropically in the trapping potential for another time of $\tau_z$. In Ref.\,\cite{rossnagel2016}, the isochoric strokes were realized by coupling the ion to a laser or external electric fields, which emulate the cold and hot baths, respectively.
\begin{figure}[t]
\centering
\includegraphics[width=15cm]{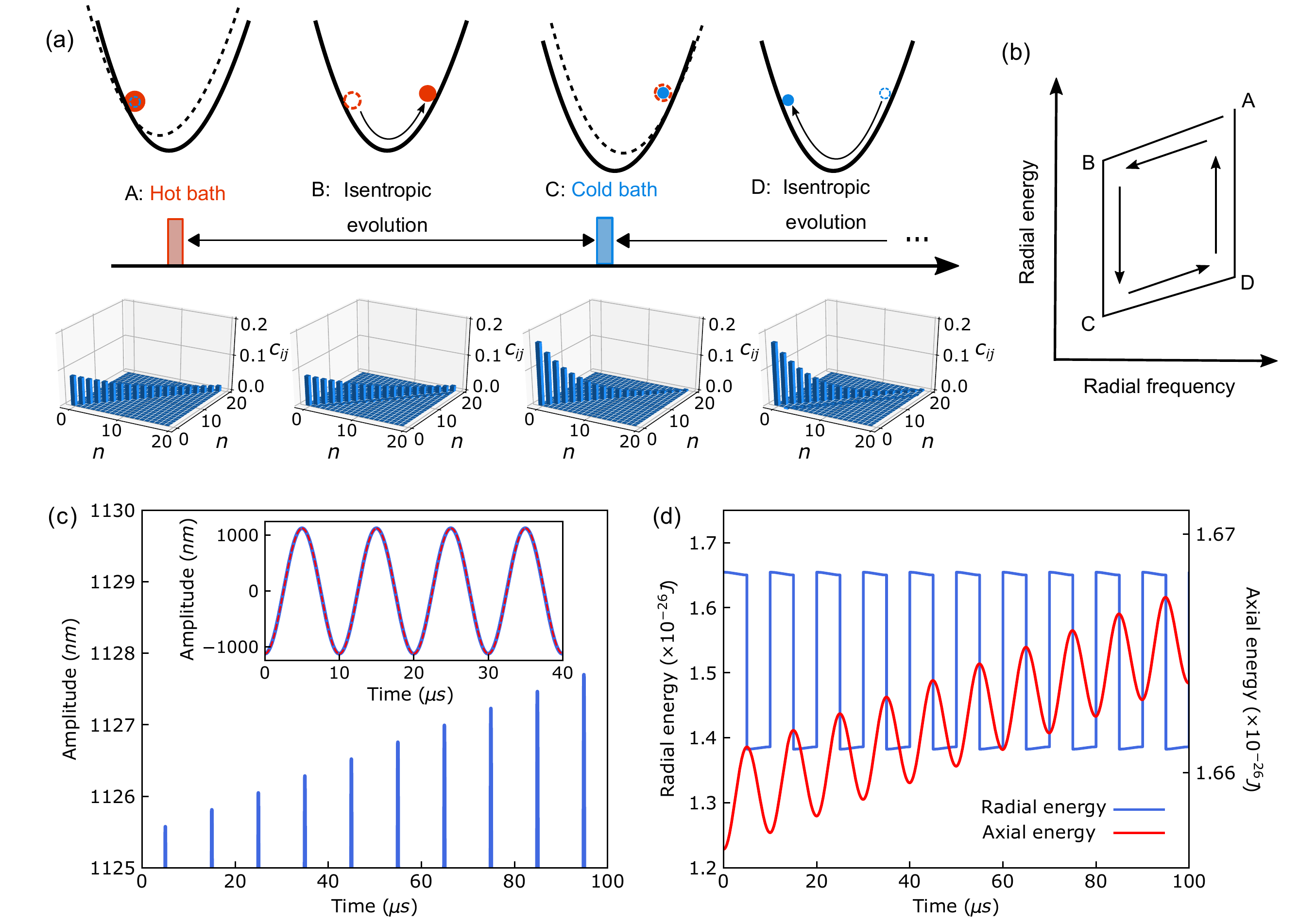}
\caption{(a) Upper: illustration of the radial and axial states of the ion that undergoes one four-stroke (A-B-C-D) cycle. Red (blue) circle represents the radial state after interaction with the hot (cold) bath. A different temperature of the radial state leads to a different amount of displacement of the axial potential. Dashed and solid black lines illustrate the axial potential before and after the bath interaction. Lower panels show the density matrix elements ($c_{ij}$) of the radial state in phonon number ($n$) basis at the end of each stroke. Only the first 20 levels are displayed. (b) Energy-frequency diagram of the radial state. The cycle is not closed due to the accumulation of energy in the flywheel. (c) Blue line shows a close-up of the axial trajectory of the ion as a function of time with $T_c=1.0$\,mK, $T_h=1.2$\,mK and $z_0=-1.1\,\mu m$. The oscillation amplitude grows linearly with the number of engine cycles. Inset: the full axial trajectory over four engine cycles. The results from numerical simulations (solid blue) and analytical calculations using Eq.\,(\ref{zn}) (dashed red) show excellent agreement. (d) Energy in the working medium (blue) and the flywheel (red) under the same condition as in (c).}
\label{Fig:enginecycle}
\end{figure}
In the following, we analyze the dynamics of the engine, combining approximated analytical solution and full numerical simulations of the exact Hamiltonian. We establish a numerical routine using a combination of propagation of the radial wavefunctions following the Liouville-von-Neumann equation\,\cite{GoerzPhD,GoerzNewtonProp} and classical trajectory simulation in the axial direction with partitioned Runge-Kutta method\,\cite{Singer2010}. This numerical platform allows us to simulate the dynamics of the engine driven by thermal and non-thermal baths. Details about this numerical routine are given in Appendix A.

We consider typical experimental parameters of a ${}^{40}\text{Ca}^{+}$ ion with $m=40$\,amu confined in a tapered Paul trap with $\theta =\pi/6$, $r_0=1$\,mm and trap frequencies of $\omega_{x0} = 2\pi\times1$\,MHz and $\omega_{z} = 2\pi\times0.1$\,MHz. Before starting the engine, all the motional states are initialized using Doppler cooling to a thermal state at 1\,mK. This results in more than 200 average phonon occupation in the axial oscillator, which justifies a classical treatment in this direction. After the initialization and at time $t=0$, the ion is at an axial position $z_0$ with axial velocity $v_{z0}=p_{z0}/m$. At this time, a hot thermal bath of temperature $\beta_h=1/k_B T_h$ is switched on. The working medium thermalizes to this temperature and has $X_0=Y_0=0, N_0=(e^{\beta_h\hbar\omega_{x0}}-1)^{-1}$. The force due to the radial-axial coupling in the small axial displacement limit becomes
\begin{eqnarray}\label{Eq:Frh}
    F_{h}=\gamma\hbar\omega_{x0}{R}_{h}\,,
\end{eqnarray}
with
\begin{eqnarray}
    {R_h}=\coth\left(\frac{\beta_h\hbar\omega_{x0}}{2}\right)\,.
\label{Eq:Rh}
\end{eqnarray}

This force leads to a displacement of the axial potential towards the open end of the taper, as depicted in the upper panel of Fig.\,\ref{Fig:enginecycle}a. After the bath interaction, the radial state expands isentropically in the potential. The classical trajectory of the ion in the $z$ direction can be solved by integrating the equations of motion\,(\ref{Eq:eom1}). After $\tau_z$, the ion reaches the position $z_1$ at $t=\tau_z$ with velocity $v_{z1}$. At this point, a cold bath with temperature $\beta_c=1/k_B T_c$ is switched on and cools the radial state to this temperature. This process is again isochoric. After cooling, the force is reduced to
\begin{eqnarray}\label{Eq:Frc}
    F_c =\gamma\hbar\omega_{x0} R_c\,,
\end{eqnarray}
where $R_c$ is given by Eq.~(\ref{Eq:Rh}) with the exchange $\beta_h\rightarrow\beta_c$.
The ion experiences a displaced axial potential towards the narrower end of the trap. 
For the next half axial oscillation period, the ion evolves again isentropically in the displaced potential.
At the end of a completed engine cycle $t=2\tau_z$, the flywheel does not restore the original $(z_0,v_{z0})$ point but ends at $(z_2,v_{z2})$ due to the work done by the forces $F_h$ and $F_c$. The energy-frequency diagram for the radial states undergoing one engine cycle is illustrated in Fig.\,\ref{Fig:enginecycle}b.

In Figs.\,\ref{Fig:enginecycle}c\,,d we present the results of numerical simulations with $T_h=1.2$\,mK and $T_c=1.0$\,mK. The blue line in Fig.\ref{Fig:enginecycle}c shows a close-up of the axial trajectory of the ion as a function of time. The oscillation amplitude grows linearly with the number of engine cycles. The inset displays the full axial trajectory over four engine cycles, where the blue line shows the numerical simulation of the exact Hamiltonian in Eq.(\ref{Eq:H2}) and the dashed red line represents the solution of using the approximated analytical expressions of $F_{h,c}$ given by Eqs. (\ref{Eq:Frh}) and (\ref{Eq:Frc}). The analytical and numerical results show excellent agreement. Figure\,\ref{Fig:enginecycle}d displays the evolution of the energy in the flywheel (red curve) and the working medium (blue curve). As a result of the linear growth of the axial oscillation amplitude, the energy in the flywheel increases quadratically with the number of engine cycles.

\section{Singe-ion heat engine as a sensitive thermal probe}
\label{sec:sensor}
In contrast to Ref.\,\cite{rossnagel2016}, where laser cooling of the flywheel was applied to reach closed-cycle operations, the engine presented here operates in open cycles where the work is stored in the flywheel. By virtue of this cyclic cumulative nature, small temperature differences can be translated to a macroscopic oscillation in the flywheel. Here, we propose using this engine as a probe for temperature differences between thermal baths. In the following, we introduce the measurement protocol and evaluate its sensitivity based on realistic experimental parameters.
\begin{figure}[t]
\centering
\includegraphics[width=15.5cm]{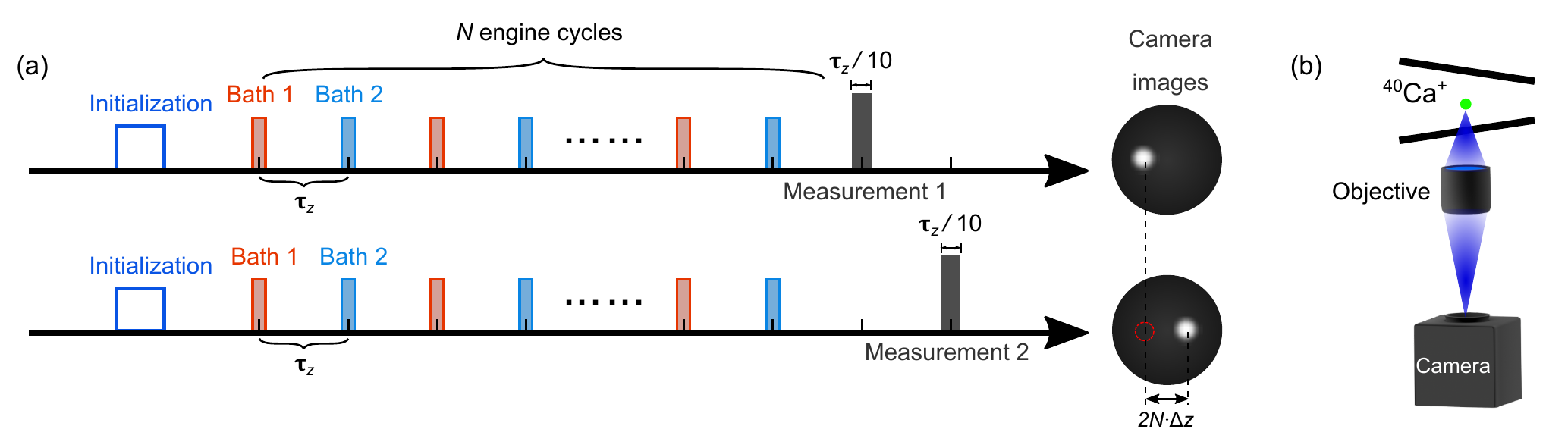}
\caption{(a) Measurement protocol. After initialization with Doppler cooling, the engine is set to operate under the driving of the two baths. After a number of $N$ engine cycles, measurements are performed to determine the extreme of the axial location. The measurements are realized by illuminating the ion with a short laser pulse of duration $\tau_z/10$, at times $\tau_z$ and $2\tau_z$ after the $2N$-th bath interaction, respectively. The physical separation of the ion's image on the camera directly translates to the axial oscillation amplitude $2N\cdot\Delta z$. It is necessary to repeat the protocol many times in order to precisely determine the amplitude. (b) The photons emitted by the ${}^{40}\text{Ca}^{+}$ ion at 397\,nm are collected by an objective and focused on a camera.}
\label{Fig:protocol}
\end{figure}
\subsection{Temperature difference estimation}
The measurement protocol is illustrated in Fig.\,\ref{Fig:protocol}. To begin with, all the motional states are initialized by Doppler cooling to a thermal distribution of temperature $T_0=1.0$\,mK. The initial axial position $z_0$ and velocity $v_{z0}$ of the ion  follow
the Boltzmann distribution  $f(z_0,v_{z0})\propto\exp[-m(v_{z0}^2+\omega_z^2z_0^2)/2k_BT_0]$.
After the initialization, the engine is set to the four-stroke operation driven by two baths of temperature $T_{1}$ and $T_2$, respectively. The position of the ion upon its $n$-th interaction with the baths can be obtained by integrating the equations of motion in Eq.~(\ref{Eq:eom1}) stroboscopically. In the small axial displacement limit ($z\ll r_0$), we arrive at
\beqar
  z_n &=&  z_0+\frac{n\hbar\omega_{x0}\gamma (R_2-R_1)}{m\omega_z^2} \quad \text{for} \quad n=0,2,4...
  \\ \nonumber
  z_n &=&  -z_0+\frac{2\hbar\omega_{x0}\gamma R_1-(n-1)\hbar\omega_{x0}\gamma (R_2-R_1)}{m\omega_z^2} \quad \text{for} \quad  n=1,3,5...,
\label{zn}
\eeqar
where $R_1$ and $R_2$ are given by Eq. (\ref{Eq:Rh}) with the corresponding temperatures $T_{1,2}$.
The change in the axial oscillation amplitude of two consecutive peaks (see  Fig.\,\ref{Fig:enginecycle}c)
\begin{equation}
\Delta z=z_{n+2}-z_n=\frac{2\hbar\omega_{x0}\gamma}{m\omega_z^2}(R_2-R_1),
\end{equation}
which in the limit of $\hbar\omega_{x0}\ll k_BT$ simplifies to
\begin{equation}
\label{eq:delta_z_approx}
    \Delta z=\frac{4 k_B\gamma}{m\omega_z^2}\Delta T,
\end{equation}
where $\Delta T=T_2-T_1$.

Note that $\Delta z$ carries the same sign with $T_2-T_1$. After $N (N\gg z_0/\Delta z)$ engine cycles, the energy fed by the engine dominates over the initial energy in the flywheel. The axial oscillation will thus be brought in phase with the bath interactions. The ion travels to the open side of the tapered potential ($z>0$) when put in contact with the bath of lower temperature, and vice versa. The temperature difference between the two baths can be determined by measuring the axial oscillation amplitude. We propose two sets of triggered measurements as follows. The first set of measurements is performed by illuminating the ion with a short laser pulse of duration $\tau_z/10$ and at a time $\tau_z$ after the $2N$-th bath interaction (see upper panel of Fig.\,\ref{Fig:protocol}a). Fluorescent light from the ion is collected by a microscope objective and imaged on a camera (see Fig.\,\ref{Fig:protocol}b). The duration of the laser pulse is chosen to be much shorter than the axial oscillation period to minimally disturb the position of the ion. To obtain a significant signal-to-noise ratio, the measurement has to be reproduced $M$ times with a restart of the engine each time. The mean position of the ion is determined by performing single-particle localization analysis\,\cite{Furstenberg2013} on the integrated fluorescent image. We note that each restart of the engine samples a random $z_0$ and $v_{z0}$ from the Boltzmann distribution. As shown in Fig.\,\ref{Fig:sensitivity}a, the influence of the distribution of $z_0$ averages out due to a large number of sampling. The measured mean axial position of the ion is thus
\beqar
  \overline{z}_{2N} = \frac{2N\hbar\omega_{x0}\gamma (R_2-R_1)}{m\omega_z^2}.
\eeqar
The second set of measurements is performed following the same procedure but at a time $2\tau_z$ after the $2N$-th bath interaction, while the $(2N+1)$-th bath interaction is skipped (see lower panel of Fig.\,\ref{Fig:protocol}a). The measured mean axial position is then
\beqar
 \overline{z}_{2N}' = -\frac{2N\hbar\omega_{x0}\gamma (R_2-R_1)}{m\omega_z^2},
\eeqar
leading to the difference between the two positions
\beqar\label{Eq:amplitude}
  \overline{z}_{2N}-\overline{z}_{2N}' = 2N\Delta z
\eeqar
a linear function of the temperature differences $\Delta T$.

Figure\,\ref{Fig:sensitivity}a displays a simulated distribution of the axial positions of the ion upon the two sets of measurements with $\Delta T=0.1$\,mK, $T_0=1.0$\,mK, $N=10^5$ and $M=2\times10^5$. Figure~\ref{Fig:sensitivity}b shows the linear dependence of $\overline{z}_{2N}-\overline{z}_{2N}'$ with the temperature difference $\Delta T$ for $N=10^5$. The black line displays the results of the approximated analytic solution. The green circles and red crosses display the same value deduced from numerical simulations performed at $T_c=1.0$\,mK and 0.2\,mK, respectively. The good agreement of the numerical simulations confirms the validity of Eq.\,(\ref{eq:delta_z_approx}) at different base temperatures.

\begin{figure}[t]
\centering
\includegraphics[width=13.5cm]{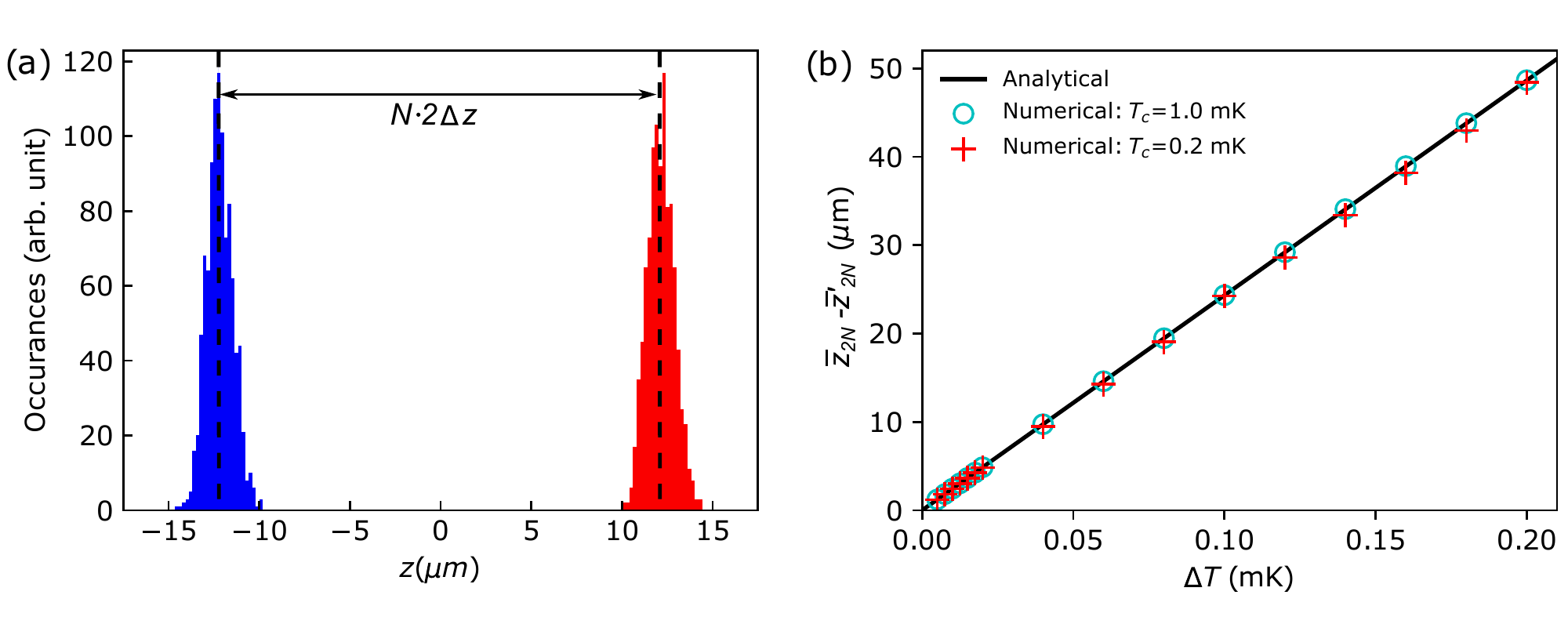}
\caption{(a) Distribution of the axial positions of the ion after $N=10^5$ number of cycles for the two sets of measurements. $M=2\times10^5$ initial positions were drafted randomly from the Boltzmann distribution with temperature $T_0=1$\,mK. (b) The oscillation amplitude $\overline{z}_{2N}-\overline{z}_{2N}'$ after $N=10^5$ engine cycles versus temperature difference $\Delta T$ between the two baths. Green circles and red crosses represent the results obtained from numerical simulations with $T_c=1.0$\,mK, 0.2\,mK, respectively. The solid black line shows the prediction of Eq.\,(\ref{Eq:amplitude}).}
\label{Fig:sensitivity}
\end{figure}

The precision in determining $\Delta T$ is limited by experimental uncertainties in measuring the amplitude $2N\Delta z$. In single-particle localization analysis, the uncertainty in the center position is given by the signal-to-noise ratio of the camera image. This is determined by the illumination time of the laser, the efficiency of photon collection, the quantum efficiency, and the background noise of the camera. In Ref.\,\cite{rossnagel2016}, an objective with a numerical aperture of 0.26 was used to collect the fluorescent photons at 397\,nm and to form an image on an intensified charge-coupled-device sensor. At each oscillation phase, $2\times10^5$ repeated measurements were necessary to obtain a localization precision of $\pm250\,$nm\,\cite{RossnagelPhD}. Assuming the same measurement conditions and using Eqs.\,(\ref{eq:delta_z_approx}) and (\ref{Eq:amplitude}), we arrive at an uncertainty of $\pm2\,\mu$K in determining $\Delta T$.

We note that the outcome of the measurement only depends on the temperature difference between the two baths and is independent of the absolute temperature in the limit of $\hbar\omega_x\ll k_BT_{1,2}$. The protocol is particularly suitable for detecting small temperature differences at high base temperatures. On the other hand, when using a bath of well-characterized temperature as a reference, the absolute temperature of an unknown bath can be determined.

\subsection{The squeezed engine: enhancing sensitivity using quantum resources}\label{sec:SqueezedAmp}
The amplitude of the measured signal can be amplified by exploiting quantum resources such as squeezing of the working medium. Here, we show that by squeezing the working medium after the isochoric strokes, a significant amplification of the oscillation amplitude can be obtained. Using this method, temperature differences smaller than $\pm2\,\mu$K might be detectable in experiments.

We assume a situation in which no prior knowledge of the baths is present. The squeezing operations described by the operator $\hat S(\xi)=\exp\big(\frac{1}{2}(\xi^*\hat a^2-\xi\hat a^{\dagger 2})\big)$ are applied to the radial state right after each of the two isochoric strokes. After squeezing, the state of the working medium is described by a squeezed thermal state, $\hat S(\xi)\rho_{th}\hat S^{\dagger}(\xi)$. 
\begin{figure}
    \centering
    \includegraphics[width=9cm]{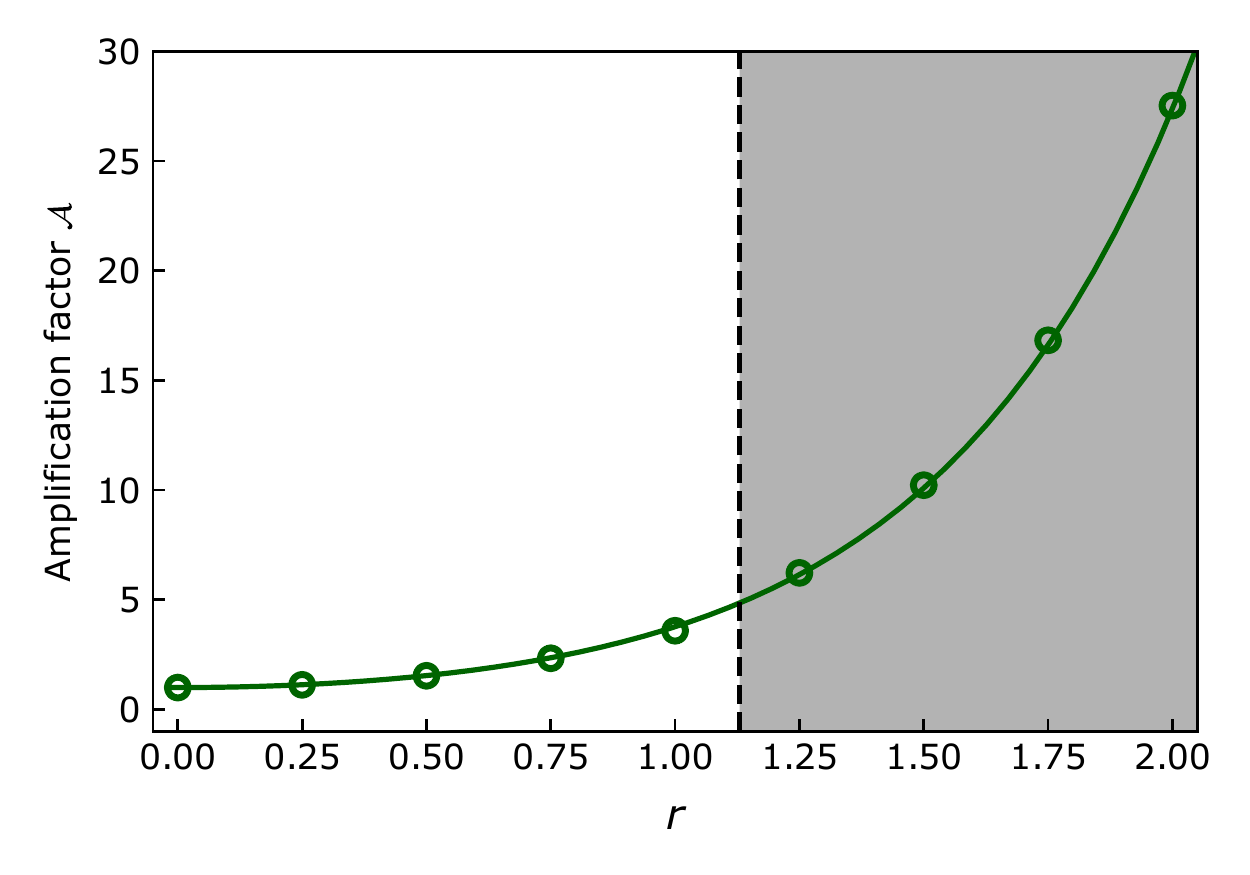}
    \caption{Amplification factor $\mathcal{A}$ as a function of the amplitude of squeezing $r$. Green circles show the results obtained from the numerical simulations. The solid line represents the outcome of Eq.\,(\ref{eq:ampBothSide}). The dashed black line indicates $r=1.13$ and the shaded area denote the region where the squeezing operations bring the working medium into the quantum regime.}
\label{fig:squeezing}
\end{figure}
In this case,  as $X_0\neq 0, Y_0\neq 0$, $R(t)$ becomes time-dependent (see Eq. (\ref{eq:second_moment})). The growth of the axial oscillation amplitude between two consecutive engine cycles becomes
\beq
\Delta z' = \frac{4\kappa\gamma\hbar}{m\omega_z}\left(\cosh(2 r)+\sinh(2 r)\cos(\alpha)/(4\kappa^2-1) \right)(n_{th}^{\rm 1}-n_{th}^{\rm 2}),
\eeq
with $\kappa=\omega_{x0}/\omega_z$, $n^{1,2}_{th}=(e^{\beta_{1,2}\hbar\omega_{x0}} -1)^{-1}$, $r$ and $\alpha$ the amplitude and phase of the squeezing parameter following $\xi=re^{i\alpha}$. Note that $\Delta z'$ is maximized when $\alpha=0$. Considering the limit of $\hbar \omega_{x0}\ll k_{B}T$, we obtain
\beq
\Delta z' =  \frac{4\gamma\hbar}{m\omega_z^2}\left(\cosh(2 r)+\sinh(2 r)/(4\kappa^2-1)\right)\Delta T.
\eeq
To evaluate the impact of squeezing the working medium on the temperature difference resolution, we define the squeezing amplification factor
\beq
\mathcal{A}=\frac{\Delta z'}{\Delta z}=\cosh(2 r)+\sinh(2 r)/(4\kappa^2-1),\label{eq:ampBothSide}
\eeq
where $\Delta z$ denotes the growth of the amplitude without squeezing. 
The dependence of the amplification factor on the squeezing parameter is depicted in Fig.\,\ref{fig:squeezing}, where the system was numerically simulated for different squeezing parameters with $T_1=0.11\,$mK, 
$T_2=0.1\,$mK and $\kappa=10$. The squeezing operations amplify the oscillation growth by an order of magnitude favoring the detection of small temperature differences. More details about the numerical simulations are presented in Appendix~B.

The squeezing operation is considered a quantum resource when the variance of one of the quadratures is smaller than $1/4$, which implies that there is a phase in which the Glauber-Sudarshan distribution turns negative \,\cite{Sudarshan1963, walls1983squeezed}. This condition is given by 
\beq
\frac{1}{2}(2 n_{th} + 1)e^{\pm 2r} < \frac{1}{4}.\label{eq:GS}
\eeq
In the example considered in Fig.\,\ref{fig:squeezing}, this corresponds to a value of $r>1.13$, as indicated by the vertical dashed line and the shaded region.
Such squeezing operations on trapped ions can be performed by fast trap voltage switching\,\cite{Heinzen1990,Alonso_2013,Burd2019}.

One should note that the optimal performance of a squeezed Otto cycle is obtained when squeezing is performed after the interaction of the working medium with the hot bath\,\cite{rossnagel2014}. In this case, the amplitude growth is typically greater than applying squeezing after both isochoric strokes, enhancing the amplitude resolution of the flywheel. Although the simple linear relation $\Delta z \propto \Delta T $ will not be satisfied, this scheme can be applied to evaluate the hot bath temperature given that the temperature of the cold bath is known.

\section{Discussions and outlook}
\label{sec:discussions}
To summarize, we proposed using an open-cycle heat engine as a sensor for temperature differences between thermal baths. Starting from a quantitative model of the heat engine, we estimated it to be capable of measuring temperature differences as small as $2\,\mu$K. Further enhancement of the signal can be achieved utilizing quantum resources such as applying squeezing operations on the working medium. Our scheme only requires initializing the engine by Doppler cooling, thus avoiding quantum state initialization processes. In the limit of $\hbar\omega_{x0}\ll k_BT$, the measurable of this scheme is independent of the base temperature of the baths, making it particularly adept at detecting small differences at a high background temperature.
When one of the baths has a well-characterized temperature, the absolute temperature of an unknown bath can be determined.

The scheme can be further extended to characterize non-thermal baths, where physical properties other than temperature can be measured in a similar fashion. Note that the bath property which governs the dynamics of the engine is imprinted on $R(t)$ (see Eq.(\ref{eq:second_moment})) of the working medium. If a bath, for example, leads to displacement or squeezing of the working medium, the relevant displacement or squeezing parameter can be evaluated quantitatively. In this respect, the indirect measurement feature at the core of this scheme avoids projective measurements on the bath and thus preserves its quantum features.

The proposed scheme is applicable to baths which can be integrated with the ion-trap platform\,\cite{Leibfried2003}.
This is so far limited to emulated baths such as lasers or external electric fields\,\cite{Myatt2000,Turchette2000}.
Nevertheless, the scheme may be of immediate interest for applications such as optimization of laser cooling routines\,\cite{Machnes2010,Lindvall2012}. In the next step, we plan to extend the proposed scheme to more realistic thermal ensembles, such as ion Coulomb crystals\,\cite{Drewsen2015} or clouds of cold neutral atoms\,\cite{Feldker2020}. By adapting the heat engine to micro-segmented or surface ion traps\,\cite{Cho2015}, the proposed scheme may also be utilized to probe local heating of ion motion due to neighbouring surfaces.

\section*{Acknowledgement}
We acknowledge financial support from  the German
Science Foundation (DFG) under project Thermal Machines in the Quantum World (FOR\,2724). E.T. acknowledges support from Project PGC2018-094792-B-I00 (MCIU/AEI/FEDER,UE), CSIC Research Platform PTI-001 and CAM/FEDER Project No. S2018/TCS-4342 (QUITEMAD-CM). We thank Samuel T. Dawkins, Daniel Basilewitsch and Daniel M. Reich for valuable discussions.

\appendix
\section{Numerical methods}\label{sec:NumericalMethods}

We use a combination of classical and quantum mechanical simulations to verify the analytical formulas of Sect.\,\ref{sec:AnalyticModel}. In order to describe the dynamics of an unbounded quantum harmonic oscillator, a truncation of the Hilbert space is necessary \cite{Fehske2009}. This truncation has to be chosen carefully to avoid reflections at the boundaries.

\subsection{Str\"omer-Verlet method}
We employ the Str\"omer-Verlet method\,\cite{Singer2010} for propagating the classical oscillator in the axial direction. The classical oscillator is described by the Hamiltonian $H(z,v)=\dfrac{v^2}{2m}+\Phi(z)$, where $\Phi(z)$ denotes a potential which is solely dependent on the position of the particle. The canonical equations of motion can be written as $z_{n+1}-2z_n+z_{n+1}=-\Delta t (q/m)\partial_z \Phi(z)$ and $v_n=(z_{n+1}-x_{n-1})/2 \Delta t$, with the time step $\Delta t=t_{n+1}-t_n$. The following recursion relations are used\,\cite{Singer2010}
\begin{align}
v_{n+\frac{1}{2}}&= v_n+\frac{\Delta t}{2}\frac{q}{m}\partial_z \Phi(z_n,t_n)\label{eqVhalf}\\
z_{n+1}&=z_n+\Delta t v_{n+\frac{1}{2}}\label{eqXp1}\\
z_{n+1}&=v_{n+\frac{1}{2}}+\frac{\Delta t}{2}\frac{q}{m}\partial_z \Phi(z_n,t_n)\label{eqVp1}.
\end{align}
\subsection{Newton propagator}
The time evolution of the radial quantum oscillator is described by the Liouville-von Neumann equation
\begin{equation}
    \mathcal{L}\left[\hat{\rho}(t)\right]=\dv{\hat{\rho}(t)}{t}=-\frac{i}{\hbar}\comm{\hH(t)}{\hat{\rho}(t)}+i\mathcal{L}_D\left[\hat{\rho}\right],
\end{equation}
with the Liouvillian $\mathcal{L}$, the Hamiltonian $\hat{H}(t)$ and the Lindbladian $\mathcal{L}_D$. The Liouvillian is expanded in Newton polynomials in order to numerically integrate the Liovulle-von Neuman equation. An arbitrary function $f(x)$ with $x\in \mathbb{C}$ can be represented in terms of Newton polynomials $R_n(x)$ with a set of sampling points $\left\{x_\ell \right\}$
\begin{equation}
    f(x)\approx \sum_{n=0}^{N-1}a_nR_n(x),\text{ with } R_n(x)=\prod_{\ell=0}^{n-1}(x-x_\ell).
\end{equation} 
The coefficients $a_n$ are computed by the recursion relation
\begin{equation}
    a_n=\frac{f(x_n)-\sum_{\ell=0}^{n-1}a_\ell\prod_{m}^{\ell-1}(x_n-x_m)}{\ell\prod_{m}^{\ell-1}(x_n-x_m)},
\end{equation}
which is referred to as \emph{divided differences}~\cite{Kosloff1994}. The first two coefficients are obtained by imposing the interpolation condition
\begin{equation}
a_0=f(x_0)\ \text{ and }\ a_1= f(x_1)-f(x_0).
\end{equation}
With $f(x)=e^{-ix\text{d}t}$ and $x=\mathcal{L}/\hbar$, the time evolution of the density matrix can be expressed as
\begin{equation}
    \hat{\rho}(\text{d}t)=e^{-\frac{i}{\hbar}\mathcal{L}\text{d}t}\hat{\rho}(0)\approx\sum_n^{N-1}a_n\left(\mathcal{L}-x_n\mathbb{1}\right)\hat{\rho}(0).
\end{equation}
For the simulations presented in this paper, we used the python implementation of the Newton propagator\,\cite{GoerzNewtonProp}. Details about the numerical methods can be found in Ref.\,\cite{GoerzPhD}. 

\subsection{Algorithm for the simulation of classical and quantum trajectories}
Description of the engine dynamics requires a combination of the classical and quantum equations of motion.
At every step of the propagation, the density matrix of the radial state, the position $z$
 and velocity $v$ of the axial oscillator, and the force $\mathcal{F}$ need to be updated. For each iteration, $R(t)$ is computed and the force $\mathcal{F}$ is updated according to Eq.(\ref{Eq:axialforce}). The axial oscillator is then propagated by $\Delta t$ using the Str\"omer-Verlet method. The newly obtained position $z(t+\Delta t)$ is used to update the Hamiltonian of our system described by Eq.(\ref{Eq:H2}) and the Liouvillian is computed. The Newton propagator~\cite{GoerzNewtonProp} is called and the state $\hat{\rho(t)}$ is propagated for the finite time step $\Delta t$. This finishes one step of the propagation. After $t=\frac{\pi}{\omega_z}$ the state is coupled to the hot or cold bath. We describe the thermalization process by updating the density matrix to the bath temperature in one step of propagation. This algorithm is illustrated in Fig.\,\ref{fig:FlowchartSim}.
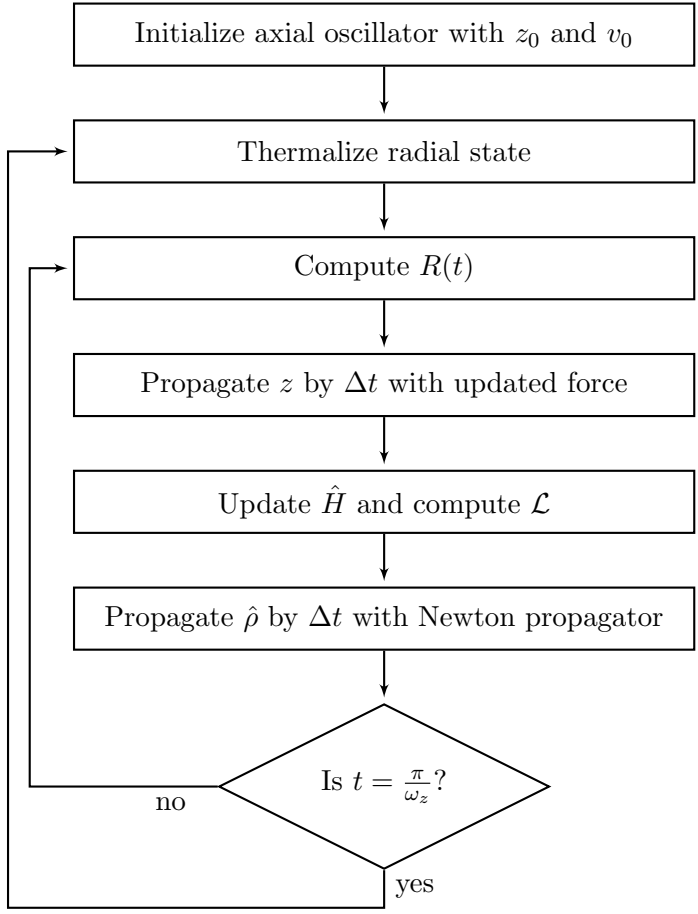
\begin{figure}[ht!]
    \centering
    \begin{tikzpicture}[font=\small,auto,decision/.style={shape aspect=2,diamond, draw=black, thick,text width=8em,align=flush center,inner sep=1pt},block/.style   ={rectangle, draw=black, thick,text width=19.3em,align=center, minimum height=2em},
line/.style    ={draw, thick, -latex',shorten >=2pt},
cloud/.style   ={draw=red, thick, ellipse,fill=red!20,minimum height=2em}]
\matrix [column sep=5mm,row sep=7mm]{
    &\node[block] (initialisation) {Initialize axial oscillator with $z_0$ and $v_0$};&\\
    &\node[block] (thermalise) {Thermalize radial state};&\\
    &\node[block] (xsqr) {Compute $R(t)$};&\\
    &\node[block] (propz) {Propagate $z$ by $\Delta t$ with updated force};&\\
    &\node[block] (updateH)  {Update $\hH$ and compute $\mathcal{L}$};&\\
    &\node[block] (evolute) {Propagate $\hat{\rho}$ by $\Delta t$ with Newton propagator};&\\
    &\node[decision] (checkTime) {Is $t=\frac{\pi}{\omega_z}$?};&\\
    };
\begin{scope}[every path/.style=line]
\path          (initialisation)     -- (thermalise);
\path          (thermalise) -- (xsqr);
\path		   (xsqr) -- (propz);
\path          (propz) -- (updateH);
\path          (updateH)   -- (evolute);
\path (evolute) -- (checkTime);
\path (checkTime.west) -- node [near start] {no} ++(-25mm,0) |- (xsqr.west);
\path          (checkTime.south)   |- node [near start] {yes}  ++(0,-5mm) -- ++(-50mm,0) |- (thermalise.west);
\end{scope}
    \end{tikzpicture}
    \caption{Flowchart of the implemented algorithm.}
    \label{fig:FlowchartSim}
\end{figure}
\newpage
\section{Squeezing the working medium}
In Fig.\,\ref{fig:squeezing} of the manuscript, we presented the amplification of the axial oscillation amplitude by applying squeezing operations on the working medium. In Fig.\,\ref{fig:App_Squeezed_Thermal_States}a the simulated axial trajectories for $r=0, 0.5$ and $1.5$ are displayed. The squeezing operations are applied after the thermalization with the cold and hot baths. For the clarity of the signal, the initial position and velocity is chosen to be zero. The corresponding density matrices after interacting with the hot bath and the squeezing operation are depicted in Figs.\,\ref{fig:App_Squeezed_Thermal_States}b-d. The phase of the entries of the density matrices $\arg(c_{ij})$ is indicated by the color scale. The applied squeezing yields a higher occupation number and excitation of the off-diagonal elements.
\begin{figure}
\centering
\includegraphics[width=17cm]{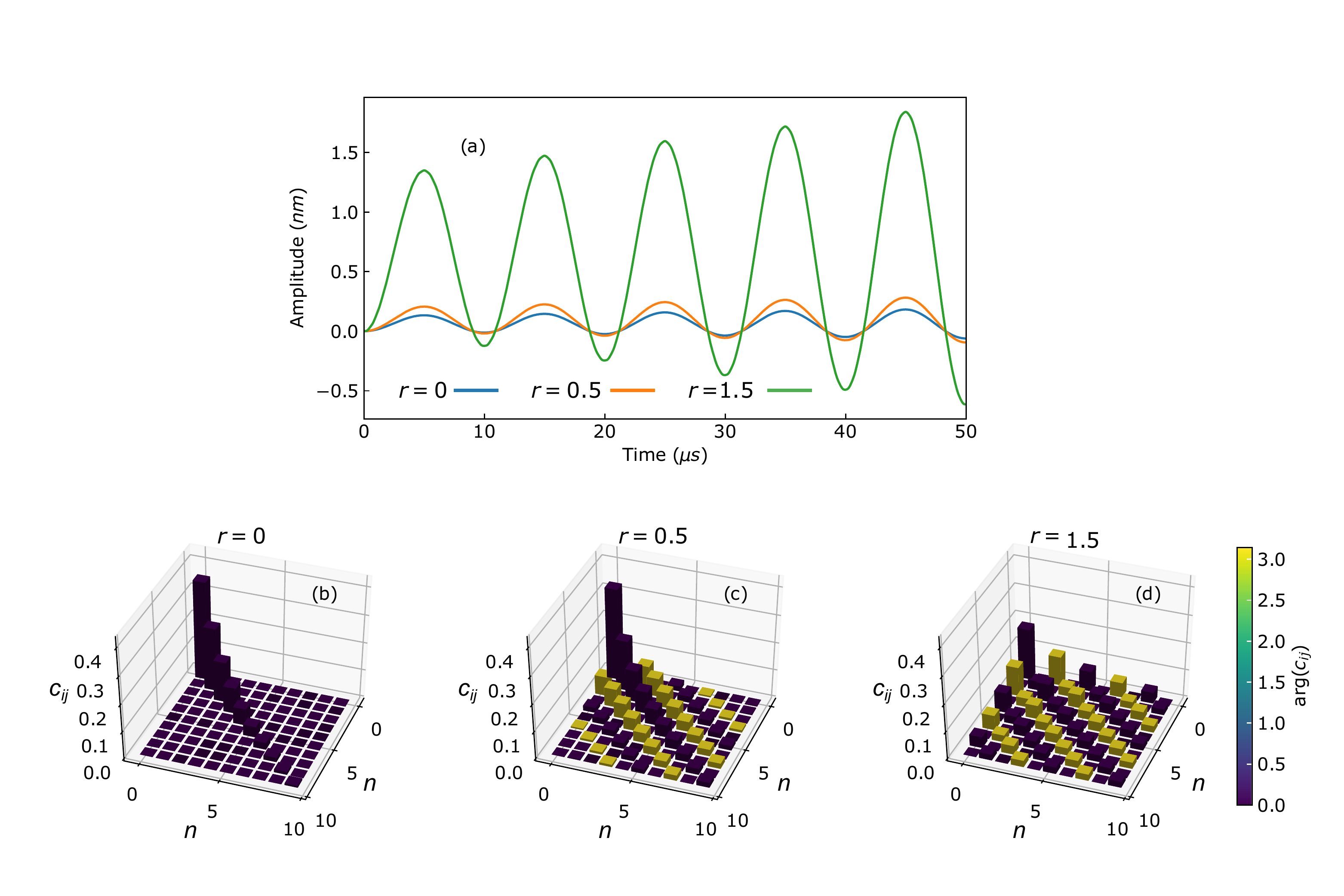}
\caption{(a) Amplification of the axial oscillation amplitude by squeezing the working medium after interaction with both baths. The simulations are performed with $T_h=0.11\,\text{mK}$ and $T_c=0.1\,\text{mK}$. The blue, orange, and green curves represent the trajectories for $r=0, 0.5$ and $1.5$, respectively. (b-d) Excerpt of the density matrices after interacting with the hot bath and the squeezing operation, where the color encodes the phase of the entries $\arg(c_{ij})$.}
\label{fig:App_Squeezed_Thermal_States}
\end{figure}
\newpage
\vspace{2cm}

\section*{References}
\bibliographystyle{iopart-num}
\bibliography{qprobe}

\end{document}